\documentclass[11pt]{article}
\usepackage[letterpaper, margin=1in]{geometry}
\usepackage[T1]{fontenc}
\usepackage[utf8]{inputenc}
\usepackage{mathptmx}
\usepackage{textcomp}
\usepackage{amsmath, amssymb}
\usepackage{booktabs}
\usepackage{array}
\usepackage{graphicx}
\usepackage{setspace}
\usepackage[hyphens]{url}
\usepackage[hidelinks, breaklinks=true]{hyperref}
\begin{document}

\begin{titlepage}
\begin{center}
\vspace*{0.8in}
{\LARGE \textbf{Lucky or Good?}}\\[6pt]
{\Large Outcome Noise, Effective Sample Size, and the Attribution of Skill}\\[36pt]
{\large Karl T. Ulrich}\\[2pt]
The Wharton School\\
University of Pennsylvania\\[2pt]
\texttt{ulrich@upenn.edu}\\[24pt]
July 30, 2026
\end{center}

\vspace{24pt}
\begin{center}\textbf{Abstract}\end{center}
\noindent When do outcome records carry enough signal to support reliable inferences about skill? When they do not, what should evaluators substitute? The framework answering the first question characterizes any decision domain with two parameters: the noise reflected in each outcome and the effective number of independent outcomes that are available over an observation window. When domains are positioned in a two-dimensional space of noise versus number of outcomes, those in which capital, prestige, and political power are routinely allocated on the basis of realized outcomes (e.g., mutual fund management, venture capital, executive performance) fall in the region where outcome records contain too little signal to support reliable individual-level inferences. Evaluating actors when outcome records are insufficient can be done by adopting the population-level empirical validation methods long used in medicine: has the actor adopted the practices that, at the population level, are associated with better outcomes?

\vspace{12pt}

\vfill
\noindent \small \textbf{Acknowledgments.} No outside funding supported this work. The paper was improved substantially by comments from Dan Levinthal and Phil Tetlock. \normalsize
\end{titlepage}

\section{Introduction}

We are confident that a chess grandmaster is genuinely better than a club player. We are less confident that a hedge fund manager who beat the market for five years is genuinely better than one who did not. These intuitive judgments reflect real differences in how much an outcome record can tell an observer about underlying skill.

This paper addresses two questions. First, under what conditions do outcome records carry enough signal to support inference about an actor's skill? Second, when conditions do not support such an inference, how can skill be assessed? The central conceptual framework for addressing the first question is a two-parameter knowability map, incorporating signal-to-noise ratio and number of observations. The methodology for evaluating skill when outcomes are insufficient borrows medicine's population-level approach to risk prediction: identify observable factors that reliably predict better outcomes across a large population, and use those factors to evaluate individuals. The main result is a prescription that the relative weight an evaluator places on outcomes versus other forms of evidence should track the position of the domain on the knowability map. Boards, allocators, founders, and individuals routinely get this backwards, weighting outcomes most heavily precisely in the domains where outcomes carry the least signal.

The challenge is not new. Knight (1921) argued that genuine profit arises from situations of true uncertainty, where the outcome distribution is not known well enough to support any return-to-skill claim from the record alone. Galton (1886) demonstrated regression to the mean, the simplest mathematical fact that threatens the inference of skill drawn from extreme observations. Tversky and Kahneman (1971, 1974) and Gilovich, Vallone, and Tversky (1985) documented that human observers reliably over-attribute outcomes to skill. Kahneman and Tversky (1973) identified the mechanism most directly at work here, base-rate neglect: evaluators over-weight distinctive, individuating evidence and under-weight the base rate. The knowability map developed below can be read as a formal statement of when the base rate should win. Fama and French (2010) and Berk and Green (2004) brought formal machinery to active portfolio management in finance. Tetlock (2005) showed that prominent political experts perform little better than chance and that the attributes predicting accuracy are not the attributes predicting reputation. Mauboussin (2012) and Frank (2016) subsequently popularized these efforts.

The organizational and strategy literature addresses similar issues with a behavioral perspective. March and March (1977) showed that careers in the Wisconsin school superintendency were indistinguishable from random processes once selection effects were accounted for. Denrell (2003) showed that the evidence of performance available to observers under-represents failure, and thus the lessons suffer survivorship bias. Denrell and Liu (2012) showed that in noisy, heavy-tailed settings the most extreme performers are disproportionately the least reliable ones. Liu and de Rond (2016) review the literature on luck in management scholarship. These concerns are visible in one of strategy's most popular genres, the success study. \emph{Good to Great} (Collins 2001) and related success studies select firms on the basis of realized outcomes and then infer strategic virtues from retrospective accounts. Rosenzweig (2007a, 2007b) showed the defect in this approach: informants' descriptions of culture, leadership, and process are themselves contaminated by knowledge of the outcomes. The framework developed here gives that critique a quantitative foundation by specifying the conditions under which the genre's inferences cannot be justified.

The behavioral and organizational literature establishes that outcome-based inference misleads in noisy settings and identifies the mechanisms. The finance lineage quantifies the problem in several consequential domains. The gap in the literature is a practical quantitative method for estimating, for any domain, when these concerns apply. Our contribution is a diagnostic for when outcome records alone are sufficient for skill attribution, and an assessment methodology for the domains where they are not. This second contribution may matter more, because it converts an impossibility result into an evaluation design problem.

For strategy the question is fundamental. The field's central phenomenon is persistent performance differences among firms and the managers who lead them. Every claim that strategy explains those differences rests on the ability to reliably detect meaningful differences. The research question also matters because attribution of performance has significant consequences. Boards fire chief executives after three years of poor financial performance, much of which reflects industry-wide shocks the executives did not cause (Bertrand and Mullainathan 2001; Jenter and Kanaan 2015). Limited partners allocate billions of dollars in capital among venture funds on the basis of the performance of three vintages, despite empirical evidence that even sophisticated institutional allocators struggle to identify true skill in this setting (Lerner, Schoar, and Wongsunwai 2007). Voters credit and blame politicians for business cycles they had limited ability to influence (Achen and Bartels 2016). If, in many of the domains where attribution matters most, the skill content of an observed track record is in theory undetectable, then a great deal of consequential behavior rests on weak epistemic foundations.

Another claim complicates the story. Even when the framework can identify an exceptional actor, apparent skill is typically contaminated by structural advantages that arose from prior outcomes, which could have been entirely the result of luck. Better deals come to top venture firms; better recruits approach top coaches; more capital flows to top managers (Sorensen 2007; Korteweg and Sorensen 2017). The framework's mean-shift detection cannot disentangle the share of an apparent effect attributable to intrinsic skill from the share attributable to the cumulative-advantage that prior performance generated. An effective evaluator must therefore make two corrections to outcome-based attribution. First, in unknowable domains, recognize that outcome records contain too little signal to support reliable individual-level inference and substitute population-validated process metrics. Second, in domains where outcomes do carry enough signal, distinguish total advantage from intrinsic skill before crediting the actor with intrinsic performance advantage.

We make an epistemic claim, not an ontological one. To say that the skill content of a record is undetectable within a working career is not to say that skill does not exist. Warren Buffett may genuinely have been more skilled than the average value investor; Reed Hastings may genuinely have been an exceptional strategist in pivoting Netflix from DVD-by-mail to streaming. The framework cannot rule out either possibility. However, it does imply that an evaluator is not warranted in inferring skill from the public record alone. This is not an indictment of any particular person, firm, or industry. The placement of a domain on the knowability map is a property of the domain, not of any individual operating in it.

The paper is organized in four remaining sections. Section 2 introduces the framework using a familiar worked example. Section 3 places six consequential decision domains on the resulting knowability map. Section 4 develops the prescriptive content for the unknowable region and notes the cumulative-advantage challenge. Section 5 concludes.

\section{When Do Outcomes Carry Signal?}

How reliably we can infer skill from a record of outcomes depends on (a) how noisy each outcome is and (b) how many independent outcomes we can observe. A single inequality using these two parameters can be used to determine when skill is statistically identifiable. Consider these three thought experiments before we derive the inequality.

A professional basketball player has shot more than 10,000 free throws by the end of their career. Each result is unambiguous; they made the shot or not. The conditions are nearly identical with each attempt. We can calculate a career free throw percentage with great accuracy. If Player A shoots 90 percent and Player B 75 percent, everyone can be confident the difference is skill.

A partner in a VC firm leads perhaps thirty investments in a career. The outcomes are unique, drawn from a heavy-tailed distribution in which one or two investments typically dominate fund returns. The elapsed time from decision to realization of outcome averages more than seven years. If their investments' internal rate of return were 15 percent and a colleague's were 8 percent, no reasonable observer can be confident, based on the outcomes alone, that the difference is skill.

A founding chief executive's record is often comprised of just one outcome: the lifetime exit value of one company (or perhaps two or three, for a serial founder). That single outcome reflects market conditions the founder did not control, competitive dynamics that played out over a decade or more, shifting macroeconomic conditions, plus a collection of decisions the founder did make. If their company sold for a billion dollars, no reasonable observer can extract from that single outcome what fraction was attributable to skill, what fraction was luck, and what fraction was the favorable weather in the market.

These three cases differ along two dimensions that taken together determine what an observer can learn from an outcome record. The first dimension is the noise structure of the domain: the dispersion, skewness, and tail behavior of outcomes conditional on a given level of true skill. Free throw outcomes are tight and well-behaved. Venture investment outcomes are wide and heavy-tailed, dominated by a small number of extreme right-tail returns. The founding chief executive's outcome is, in the limit, a single draw from an even more extreme distribution. The second dimension is the effective sample size achievable in a working career. The basketball player makes thousands of nearly independent decisions per year, each resolved within seconds. The venture capitalist makes a few investments per year, each typically resolving over many years, with substantial vintage and sector correlation. The founding chief executive produces effectively one resolved outcome over a decade, the fate of one company.

The two dimensions are largely independent. A domain can have tight noise but few decisions (a single high-stakes negotiation), wide noise and many decisions (a quantitative trader in volatile markets), wide noise and few decisions (most consequential life choices), or tight noise and many decisions (a free-throw shooter, a chess player, a sprinter). The detectability of skill is jointly determined by where the domain sits on these two axes.

The relationship among noise, sample size, and confidence in attributing outcomes to skill is captured by a single inequality. For a domain with within-actor noise in outcomes characterized by a standard deviation \(\sigma\) and an effective sample size \(N_{\text{eff}}\), the smallest skill increment \(\alpha^*\) that an observer can reliably distinguish from chance is approximately \(2.5 \, \sigma / \sqrt{N_{\text{eff}}}\) at a one-sided 5 percent significance level and 80 percent power (the conventional benchmarks in power analysis). The minimum detectable skill rises linearly with noise and falls with the square root of effective sample size. Doubling the noise therefore doubles the threshold, whereas quadrupling the effective sample size halves it.

Two definitional notes. First, \(\sigma\) is the dispersion of outcomes conditional on an actor's true skill. The pooled cross-actor dispersion is \(\sqrt{\sigma^2 + \sigma_{\text{skill}}^2}\) and the binary calibrations below use the pooled value. This overstates \(\sigma\) by a few percent at the SNR levels we consider. Second, the constant 2.5 applies to comparing one actor's record against a known benchmark. Distinguishing two actors from each other, each observed \(N_{\text{eff}}\) times, increases the constant by \(\sqrt{2}\), to approximately 3.5.

These notes generalize to a taxonomy of estimands, with the evaluator's specific question determining the exact arithmetic. Four questions recur: (1) Is this actor better than average? That is the one-sample detection problem for which the constant 2.5 applies. (2) Is actor A better than actor B? That is the two-sample problem, with the constant inflated by \(\sqrt{2}\). (3) How will this actor perform in the next period? Prediction is more forgiving than detection in one respect, because even an undetectably small skill signal shifts the optimal forecast a little, but the weight the forecast places on the record is the same reliability quantity. So, in unknowable domains the record barely moves the prediction. (4) Who is the best actor in the population? Selection is the hardest of the four questions. The winner of a ranking is disproportionately benefitted by favorable noise. Therefore, the apparent gap between the selected actor and the rest overstates the true gap, and the multiplicity correction discussed in Section 3 applies. The knowability map in this paper, Figure 1, is drawn for the first question. Questions 2 and 4 are strictly harder, and question 3 is governed by the same reliability that makes detection fail, so the map errs on the side of detectability. Placements into the unknowable region only deepen for questions 2 through 4.

Normalizing by the cross-actor standard deviation of true skill in the domain (\(\sigma_{\text{skill}}\)) gives the dimensionless per-observation signal-to-noise ratio \(\text{SNR} = \sigma_{\text{skill}} / \sigma\), and the detectability inequality takes the compact form \(\text{SNR} \times \sqrt{N_{\text{eff}}} > 2.5/d\), where \(d\) is Cohen's standardized effect size for the skill difference we wish to detect (Cohen 1988). The left-hand side is the actor's ``detection budget.'' Detecting a medium effect (\(d = 0.5\), i.e., 0.5\(\sigma_{\text{skill}}\) ) requires the budget to exceed 5.0; detecting a large effect (\(d = 0.8\)) requires it to exceed 3.13. We standardize deliberately by \(\sigma_{\text{skill}}\) rather than by the outcome noise that anchors Cohen's within-group convention: \(d = 0.5\) here means two actors half a skill-SD apart in the latent skill distribution, even though the corresponding per-observation effect in outcome units, \(d \times \text{SNR}\), falls below Cohen's small benchmark in every domain we consider.

To solidify the metric, we work out a simple example drawn from a single domain that nearly every reader has intuitions about, professional basketball in the United States. The four instances differ from one another only in their parameters. They share the same league, the same actors, the same physical reality, the same officials. The framework predicts substantially different inferential possibilities across the four instances, reflecting differences in \(\sigma\), \(\sigma_{\text{skill}}\), and \(N_{\text{eff}}\) alone.

\textbf{Instance one: a player's career free throw record.} Per attempt, the outcome is Bernoulli with \(\sigma \approx 0.43\) around the league mean of 0.75. Cross-player skill SD is \(\sigma_{\text{skill}} \approx 0.10\), so SNR = \(\sigma_{\text{skill}}/\sigma \approx 0.23\). A career player attempts thousands of nearly independent free throws with instant feedback, so \(N_{\text{eff}} \approx 5{,}000\). The detection budget SNR \(\times \sqrt{N_{\text{eff}}} \approx 16\), well above the medium-effect threshold of 5.0. Skill differences of one or two percentage points can be reliably detected from career records, and a skill difference of \(d = 0.5\) can be detected in approximately 470 free throws, which is less than a single NBA season.

\textbf{Instance two: a head coach's single regular-season win percentage.} Per game, \(\sigma \approx 0.5\) (the Bernoulli maximum). Cross-coach skill SD on team win percentage is modest, \(\sigma_{\text{skill}} \approx 0.03\), so SNR \(\approx 0.06\). A regular NBA schedule is 82 games. The detection budget is \(0.06 \times \sqrt{82} \approx 0.54\), an order of magnitude below the medium-effect threshold of 5.0 and well below the large-effect threshold of 3.13. The framework correctly predicts that the win-loss record of a single season tells the observer essentially nothing about coaching skill that they did not know at the start of the season. This contradicts the annual ritual of coach-of-the-year voting and mid-season firings.

\textbf{Instance three: a head coach's career win percentage over a thousand games.} With \(N\) raised to 1,000 over a long career, the detection budget is \(0.06 \times \sqrt{1{,}000} \approx 1.9\). The budget remains below the medium-effect threshold but is now sufficient to detect exceptionally large effects. This is the range in which a small set of coaches become identifiable even to non-fans; for example, Pat Riley, Phil Jackson, Gregg Popovich, Mike Krzyzewski, and Bob Knight. Their detectability arises from unusual career length combined with consistent over-performance, not a property of typical coaching.

\textbf{Instance four: the championship as evidence that a team is the best.} A single championship contains far less information than most people assume. Even the team with the highest true skill wins the title in any given year only with probability roughly 0.20 to 0.25, because best-of-seven series go the favorite's way only about 71 percent of the time and four such series occur in sequence. A single championship updates the posterior probability of ``this team is genuinely the best'' from the prior of roughly 1/30 only to about 20 to 25 percent. Winning a title is strong evidence relative to the prior, an update by a factor of six or more, but still leaves it more likely than not that some other team was better.

The four instances illustrate the framework's central claim. The same domain, the same actors, the same physical reality, can support reliable skill inference at the parts-per-thousand level when SNR and \(N_{\text{eff}}\) are favorable, and provide essentially no inferential traction when they are not. Treating NBA championships and single-season coaching records as definitive evidence of skill therefore involves an attribution that the outcomes alone cannot analytically support. The rest of the paper applies this same logic to domains where the stakes may be larger.

A note on the style of the inequality. Specialists may attack the luck-versus-skill problem with more elaborate tools: false discovery rate methods that control for the multiplicity of ex post selected star performers (Barras, Scaillet, and Wermers 2010), hierarchical Bayes models that shrink individual estimates toward the population (Korteweg and Sorensen 2017), and structural decompositions of persistence into skill and matching (Sorensen 2007). That machinery produces sharper answers within a domain, at the price of domain-specific data and modeling. The two-parameter reduction deliberately trades that precision for portability. Any evaluator who can approximate a noise level and an effective sample size can place their domain on the map and read off whether the inference they are contemplating is even a candidate for support. The inequality is a screening instrument, not a substitute for domain-specific econometrics that will be used repeatedly in the same setting. When the inequality signals marginal detectability, the heavier tools may be the right next step.

\section{A Knowability Map for Economic Attribution}

Plotting SNR on the horizontal axis and \(N_{\text{eff}}\) on the vertical axis, both on log scales, gives a coordinate space within which any decision domain can be placed. Recall that the detection budget from Section 2 is SNR \(\times \sqrt{N_{\text{eff}}}\). Iso-detectability contours run diagonally with slope \(-2\) (in log-log coordinates). We classify domains as \emph{knowable} if their detection budget exceeds 5.0 (medium effects detectable), \emph{detectable with discipline} if the budget falls between 3.13 and 5.0 (only large effects detectable), and \emph{essentially unknowable} if the budget falls below 3.13 (even large effects are not reliably distinguishable from chance).

Figure 1 positions ten decision domains on the resulting map. The choice of domains focuses on the economic and policy cases where attribution claims carry the largest consequence (e.g., active mutual fund management, venture capital partnership, Fortune 500 chief executive performance, founder performance, Federal Reserve forecasting, film direction), with high-frequency trading at the deeply knowable extreme and three sports or game examples (NBA player free throws, tournament chess, Tetlock-style geopolitical superforecasting) as anchors for intuition. Each placement carries an uncertainty area constructed from the calibration ranges reported below. The paper's claims are supported by the approximate region of the placement and not by the point estimates.

\begin{center}
\includegraphics[width=0.85\textwidth]{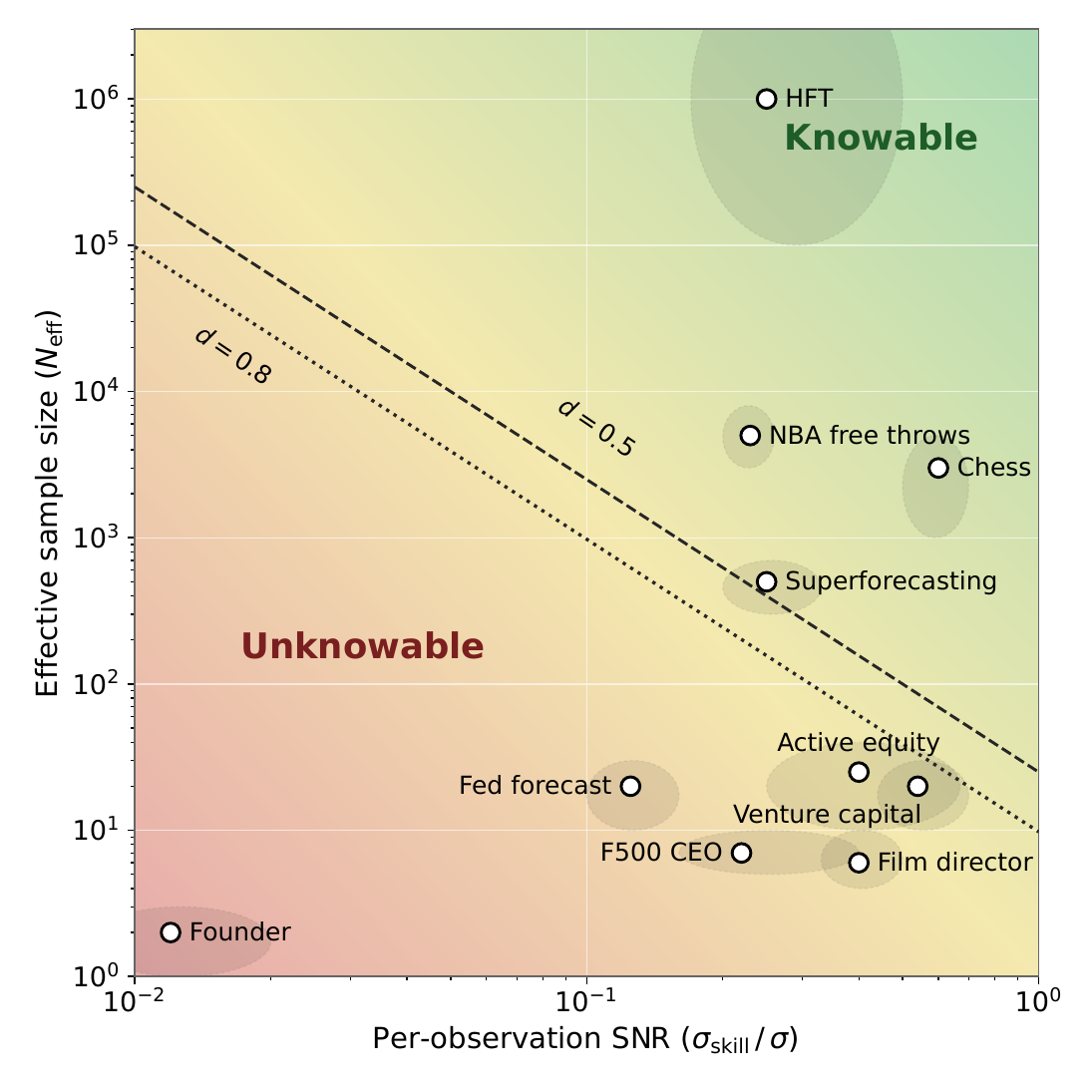}

\textit{Figure 1. The knowability map. Ten decision domains placed on the (SNR, $N_{\text{eff}}$) plane. Dashed contours mark the d = 0.5 (medium effect) and d = 0.8 (large effect) detection thresholds. Shaded ellipses show the calibration ranges reported in the text. The essentially unknowable verdicts hold across the ranges except at the most favorable corners of the active equity and venture capital regions, where the budget approaches the large-effect threshold, consistent with the top-tail qualifications in the text.}
\end{center}

\textbf{NBA player career free throws} (knowable). \(\sigma \approx 0.43\), \(\sigma_{\text{skill}} \approx 0.10\), SNR \(\approx 0.23\), \(N_{\text{eff}} \approx 5{,}000\), detection budget \(\approx 16\). The cleanest knowable case in the paper, anchored in the empirical NBA distribution of career free-throw rates.

\textbf{High-frequency systematic trading} (knowable). \(\sigma \in [0.01, 0.03]\) per trade, \(\sigma_{\text{skill}} \approx 0.005\) (about 50 basis points), SNR \(\approx 0.25\), \(N_{\text{eff}} \in [10^5, 10^7]\), detection budget at \(N = 10^6\) approximately 250. The deepest position in the knowable region. The cross-firm skill SD is a rough approximation because the industry is famously secretive about per-trade economics, but the location on the map is robust to any plausible \(\sigma_{\text{skill}}\) in the 5 to 50 basis point range because effective \(N\) over a career is in the millions.

\textbf{Tournament chess} (knowable). Per game, the outcome is Bernoulli with \(\sigma \approx 0.5\). Cross-tournament-player SD of true skill in win-rate units is approximately \(\sigma_{\text{skill}} \approx 0.30\), corresponding to an Elo rating SD of about 250 across active rated players, so SNR \(\approx 0.60\). A serious tournament player accumulates between 1,000 and 5,000 rated games over a career with low cross-game correlation. Detection budget at \(N = 3{,}000\) is approximately 33, more than five times the medium-effect threshold. The Elo rating system (Elo 1978) is itself a formal skill-detection apparatus, and rating differences as small as 30 to 50 Elo points are routinely identified in active players. Chess serves as a familiar accessibility anchor for the framework's intuition.

\textbf{Tetlock-style geopolitical superforecasting} (boundary). \(\sigma \in [0.15, 0.25]\) per question (Brier score variance), \(\sigma_{\text{skill}} \approx 0.05\) (Mellers et al.~2014), both on the Good Judgment Project's 0-to-2 Brier convention (the SNR is invariant to the scale convention because numerator and denominator share units), SNR \(\approx 0.25\), \(N \approx 500\) over a four-year tournament, detection budget \(\approx 5.6\). The placement reads as detectable-with-discipline because the disciplining infrastructure (a tournament with question resolution and Brier scoring) is what moves the domain into detectability at all. Forecasting in its natural state, where pundits make predictions that are usually never resolved against ground truth, is essentially unknowable. The Good Judgment Project changed the data-generating process by imposing tournament structure: clearly resolvable questions, mandatory probability statements, calibration scoring, structured aggregation. The forecasters did not change. The evaluation regime did. This case suggests that domains can be moved between regions with structural intervention. Mellers et al.~(2015) provide the direct evidence: forecasters selected into the top two percent after the first tournament year defied regression to the mean in each of the following two years, exactly the pattern expected if tournament structure had moved real skill into the detectable region. The project's practice of standardizing scores within question addresses the concern that Brier outcomes conflate skill with question difficulty. Mellers et al.~also describe superforecasters as partly discovered and partly created: assignment to elite teams is itself an enriched environment generated by prior performance, a benign instance of the cumulative-advantage mechanism developed in Section 4.

\textbf{Long-only public equity active management} (essentially unknowable, top decile only). \(\sigma \in [0.03, 0.08]\) per year (Wermers 2000), \(\sigma_{\text{skill}} \approx 0.02\) per year (Wermers 2000; Fama and French 2010; Barras, Scaillet, and Wermers 2010), SNR \(\approx 0.40\), \(N_{\text{eff}} \in [10, 40]\) over a 30-year career after realistic correlation adjustment (\(\rho \in [0.05, 0.15]\) across portfolio decisions), detection budget at \(N = 25\) approximately 2.0. Even managers with true skill of 1.6 percent annual alpha (\(d = 0.8\)) cannot be reliably distinguished from luck at typical career \(N\). Manager identification at the very top tail (top decile or higher, where true alpha may exceed 3 percent annually and \(d\) exceeds 1.5) becomes marginally feasible. This is consistent with the long-running empirical finding that few mutual fund managers can be statistically distinguished from luck (Berk and Green 2004; Fama and French 2010; Barras, Scaillet, and Wermers 2010). The strategy field has its own version of this result. Denrell (2004) shows that sustained competitive advantage, the pattern the persistence literature reads as evidence of superior firm capability, arises in a pure random walk model containing no skill differences at all. That the same lesson emerges independently from mutual fund econometrics and from strategy-field simulation is itself evidence for the cross-domain portability the framework claims. Persistence alone cannot certify skill wherever noise is large relative to the skill spread.

\textbf{Venture capital partnership} (essentially unknowable, top tail only). \(\sigma \in [1.0, 1.6]\) per investment (log return SD; Cochrane 2005; Korteweg and Sorensen 2017), \(\sigma_{\text{skill}} \approx 0.7\) in log-return units (Korteweg-Sorensen variance decomposition), SNR \(\approx 0.54\), \(N_{\text{eff}} \in [10, 30]\) after correlation adjustment for vintage and sector concentration, detection budget at \(N = 20\) approximately 2.4. Medium-magnitude skill differences in venture capital cannot be reliably detected from outcomes within a career; only very-large effects (\(d > 1.5\)) become marginally identifiable. Sorensen (2007) further shows that approximately two-thirds of the apparent persistence in top-quartile VC returns reflects selection of deal flow (the better deals come to the better-known firms) rather than investment skill per se, which the framework's mean-shift treatment cannot disentangle. The top of the partner distribution is potentially identifiable as a boundary case, but even then the apparent skill is contaminated by the cumulative-advantage problem developed in Section 4.

\textbf{Non-founder Fortune 500 chief executive} (essentially unknowable). A chief executive who is not the company's founder typically serves a tenure of 5 to 10 years with on the order of 5 to 15 truly independent strategic decisions. Per-year benchmark-adjusted firm return SD is \(\sigma \in [0.10, 0.25]\). Cross-CEO SD of true effect on firm outcomes is \(\sigma_{\text{skill}} \approx 0.04\) (Bertrand and Schoar 2003; Kaplan, Klebanov, and Sorensen 2012), so SNR \(\approx 0.22\). Effective \(N\) over a tenure, after collapsing highly correlated quarterly observations into truly independent strategic decisions, is on the order of 5 to 10. Detection budget at \(N = 7\) approximately 0.58, an order of magnitude below the large-effect threshold. Even an exceptional CEO whose true effect on firm performance is 1.5 SD above the median (a six percentage point per year benchmark-adjusted contribution) cannot be reliably distinguished from a beneficiary of favorable industry conditions on outcome data alone. Bertrand and Mullainathan (2001) document directly that observed CEO performance is contaminated by industry-wide shocks.

\textbf{Federal Reserve macroeconomic forecasting} (essentially unknowable). Per-forecast noise in one-year-ahead real GDP growth predictions is on the order of \(\sigma \approx 0.04\). Cross-forecaster SD of true skill is approximately \(\sigma_{\text{skill}} \approx 0.005\), so SNR \(\approx 0.125\). A career forecaster generates perhaps 20 truly independent forecasting cycles after accounting for correlation across nearby quarters. Detection budget \(\approx 0.56\), well below the large-effect threshold. Loungani (2001) and successors document that professional consensus forecasts predicted only two of the 60 recessions in their sample a year or more in advance; the rest were recognized only as they arrived or afterward. The framework predicts that no single career produces enough independent outcomes to support reliable individual-level skill inference in macroeconomic forecasting; population-level forecasting tournaments (as in the Tetlock superforecasting case above) are the structural intervention that can move forecasting into the detectable region.

\textbf{Film director, commercial career} (essentially unknowable). A working director makes between 5 and 15 features over a career. Per-film commercial outcomes are heavy-tailed (De Vany 2003), with log-revenue SD on the order of \(\sigma \in [2.0, 3.0]\). Cross-director SD of true commercial skill in log-revenue units is approximately \(\sigma_{\text{skill}} \approx 1.0\), so SNR \(\approx 0.40\). With substantial period and genre correlation across a director's filmography, effective \(N\) after correlation adjustment is on the order of 6. Detection budget \(\approx 0.98\), well below the large-effect threshold. Salganik, Dodds, and Watts (2006) demonstrate experimentally that cultural-market outcomes are path-dependent in ways that further compromise outcome-based attribution: early random success can lock in lasting commercial success even when underlying creative quality is similar.

\textbf{Founder CEO of a single venture} (essentially unknowable, extreme). A founder typically launches one or two ventures over a career; the conditional outcome distribution is dominated by a very heavy right tail (Cochrane 2005). For a single venture, \(N_{\text{eff}} = 1\) in the relevant sense, and \(\sigma_{\text{skill}}\) is essentially undefined because the founder skill distribution is too heavy-tailed for the second moment to be reliably identified. SNR is effectively zero. The framework's verdict is that no observer can reliably distinguish founder skill from luck on the basis of a single outcome, regardless of how impressive that outcome was. Gompers, Kovner, Lerner, and Scharfstein (2010) document modest persistence in founder outcomes across multiple ventures (success rates of approximately 30 percent for previously successful founders versus 18 percent for first-time founders), which is informative at the population level but not at the level of individual attribution from a single venture. The founder case is also where the framework's parameters themselves become unidentifiable. With one outcome per actor, the observable cross-sectional variance is the sum \(\sigma^2 + \sigma_{\text{skill}}^2\), and no data the domain generates can split it into noise and skill; the detection failure and the identification failure are the same fact viewed twice. This is the strongest form of unknowability the framework recognizes.

The \(N_{\text{eff}}\) values above embed judgments about correlation across an actor's outcomes. Making the adjustment explicit, the equicorrelation model \(N_{\text{eff}} = N / (1 + (N-1)\rho)\) maps \(N\) raw outcomes with pairwise correlation \(\rho\) into their independent-information equivalent. Table 1 reports the detection budgets this implies for the consequential domains across a range of \(\rho\), holding each domain's SNR at its calibrated value. Two features stand out. First, the unknowable-region verdicts are robust: active equity, venture capital, chief executives, macroeconomic forecasting, and film direction sit below the large-effect threshold of 3.13 even at \(\rho = 0\), the most favorable possible correlation structure, and correlation only deepens the verdicts. Second, the model has a ceiling: as \(N\) grows, \(N_{\text{eff}}\) approaches \(1/\rho\), so once an actor's outcomes share material correlation, no career length rescues detectability; at \(\rho = 0.10\), no amount of experience yields more than ten independent observations. The superforecasting placement is the one that moves. Its detectability depends on question diversity holding cross-question correlation near zero, which is exactly what its tournament design works to achieve.

\begin{center}
\begin{tabular}{lcccccc}
\toprule
Domain & SNR & Raw $N$ & $\rho = 0$ & $\rho = 0.05$ & $\rho = 0.10$ & $\rho = 0.20$ \\
\midrule
Active equity & 0.40 & 30 annual returns & 2.19 & 1.40 & 1.11 & 0.84 \\
Venture capital & 0.54 & 30 investments & 2.96 & 1.89 & 1.50 & 1.13 \\
Fortune 500 CEO & 0.22 & 40 quarters & 1.39 & 0.81 & 0.63 & 0.47 \\
Fed forecast & 0.125 & 40 quarters & 0.79 & 0.46 & 0.36 & 0.27 \\
Film director & 0.40 & 10 features & 1.26 & 1.05 & 0.92 & 0.76 \\
Superforecasting & 0.25 & 500 questions & 5.59 & 1.10 & 0.78 & 0.56 \\
\bottomrule
\end{tabular}
\end{center}

\emph{Table 1. Detection budgets \(\text{SNR} \times \sqrt{N_{\text{eff}}}\) under the equicorrelation model. Raw \(N\) values are illustrative career observation counts. The medium-effect threshold is 5.0; the large-effect threshold is 3.13.}

Four comments about the map:

First, every domain we placed has SNR less than 1. The per-observation noise \(\sigma\) exceeds the cross-actor skill spread \(\sigma_{\text{skill}}\) in every consequential decision domain we consider. Detection of skill is therefore always a problem of integrating signal over many observations and not a problem of reading skill cleanly off any single observation. This is the structural reason that career-feasible \(N_{\text{eff}}\) is the binding constraint in almost every case.

Second, the placement of the three cases where the largest economic and political stakes are associated with attribution claims (e.g., long-only active equity, venture capital partnership, and non-founder chief executive performance) all sit firmly in the essentially unknowable region. Together they account for trillions of dollars of capital allocation and the bulk of executive compensation in the developed world. The framework predicts that, with the exception of the very top tails of these distributions, the practice of attributing observed performance to skill in these domains is not supportable from outcome data alone. The median active equity manager cannot be reliably distinguished from an unskilled or zero-alpha manager. The typical venture capital partner cannot be distinguished from a beneficiary of favorable deal flow. The conventional Fortune 500 chief executive cannot be distinguished from a beneficiary of favorable industry conditions. This is not a claim that these individuals lack skill. It is a claim that we cannot reliably know which actors are skilled from the public record alone, and that the population-level practice of treating outcome track records as if they identified skill rests on weaker evidentiary foundations than its practitioners assume.

Third, the binary language of knowable and unknowable is a deliberate simplification of a continuous quantity. A Bayesian evaluator never declares skill undetectable. The weight a posterior places on an actor's record relative to the prior grows with \(\text{SNR}^2 \times N_{\text{eff}}\), so the map's contours are level sets of the information content of a career record, and ``essentially unknowable'' means shrinkage so severe that the record barely moves the prior. We retain the detection framing because it matches the institutional question (fire or retain, allocate or pass) and because it connects to standard power analysis, but nothing in the argument depends on where the thresholds sit. The multiplicity problem compounds the shrinkage problem. When an actor comes to an evaluator's attention because of an extreme record, as star managers do, inference must also correct for selection across the whole population, the false-discovery logic that Barras, Scaillet, and Wermers (2010) formalize for mutual funds. The map's budgets are computed for a single actor specified in advance; ex post selection makes every inference strictly harder than the map suggests.

Fourth, what inference is warranted inside the essentially unknowable region? Here the map should be read together with Denrell and Liu (2012). Their result is adjacent to ours and in one respect stronger. In domains where noise is high and outcome distributions are heavy tailed, the most extreme observed performance is disproportionately generated by high-variance, unreliable processes, so an extreme outcome is evidence of unreliability rather than of skill. The detectability inequality gives the general condition under which outcome records can support skill inference at all. Denrell and Liu characterize the inference that remains warranted in the region where that condition fails and tails are heavy: the top of the outcome distribution should sometimes be actively discounted, not merely treated as uninformative. The ``top tail only'' qualifications attached to the active equity and venture capital placements above should therefore be read with this caveat in hand. In these settings, the top tail is not the exception to the map's pessimism; rather it is where outcome-based inference is least trustworthy.

\section{What to Do When Outcomes Do Not Carry Signal}

The descriptive map of Section 3 raises an obvious question. If outcomes in essentially unknowable domains do not allow individual skill to be determined, what should we do instead? The failure of outcome-based inference in noisy domains has been documented in several distinct literatures. To our knowledge, however, those literatures have not produced a general method for evaluating actors when outcomes are insufficient. The answer is to substitute population-validated process and structural metrics. The methodology that justifies this approach is borrowed from prediction of individual risk in medicine.

The pessimism of the framework operates at the individual-level, not the population-level. Across many actors in the same domain, the total observation count is much larger than any single actor's career \(N\). The population is large enough to support the statistical inference that an individual record cannot. We do this by estimating relationships between observable process and structural features and realized outcomes, identifying which features carry predictive content, and quantifying how much. A process metric that has been shown across hundreds or thousands of actors to correlate with realized outcomes carries inferential weight for individual evaluation, because we can use the population-estimated relationship as a surrogate for the missing individual-level outcome data. The right prescription is not just ``use process metrics'' but ``use process metrics whose predictive validity has been established on populations.''

The case for population validation is strengthened by what we know about the alternatives. The evaluator's accumulated experience is not a substitute, because experiential learning from outcome feedback is itself systematically biased. Denrell and March (2001) show that adaptive learners abandon alternatives after early failures and therefore never generate the observations that would correct an unluckily negative first impression, the hot stove effect. Denrell, Fang, and Levinthal (2004) show that when direct outcome feedback is too sparse to learn from, learners substitute feedback from their own mental models of the domain, and the learning then inherits whatever errors the models contain. An evaluator who trusts seasoned judgment over population-validated metrics in an unknowable-region domain is relying on precisely the biased learning processes these results characterize.

The methodology has the same statistical structure as the Framingham Heart Study. Framingham followed thousands of patients in Framingham, Massachusetts, beginning in 1948, recording observable features (age, blood pressure, cholesterol, smoking, diabetes) at biennial visits and tracking sparse but unambiguous cardiovascular outcomes. The relationship estimated across that population (the Framingham Risk Score) is then used to predict cardiovascular risk for new patients from their observable features alone, without waiting for the events that the prediction is trying to anticipate.

The structural parallel to the unknowable-region attribution problem is close in three respects, though imperfect for reasons we note at the end. Individual-level data is insufficient at the moment of decision: a forty-five-year-old without prior cardiovascular events has at most a sparse personal record; a Fortune 500 chief executive in year three of a tenure has perhaps five resolved strategic decisions. The population of similar units does have enough aggregate data to estimate the feature-to-outcome relationship: Framingham did this over decades for cardiovascular events, and Bertrand and Schoar (2003), Kaplan, Klebanov, and Sorensen (2012), and Cremers and Petajisto (2009) have done analogues for chief executives and active mutual funds. The population-estimated relationship is then applied at the individual level using observable features alone, without ever requiring the individual to have generated enough outcome data to be evaluated on outcomes alone. The Framingham case differs in why the individual data is insufficient (sparse outcomes versus noise-contaminated outcomes) but the methodological approach is the same.

A note on the unit of attribution before the examples. Population validation can be applied at more than one level: to an individual actor, to a team or program, or to a firm-level process. The appropriate level of analysis matters because process-level and individual-level evidence support different decisions. Process evaluation supports investment in the process, while individual attribution supports promotion, retention, and pay. The four examples below span these levels.

\textbf{Active share in mutual fund management.} Active share is the fraction of a fund's portfolio holdings that differs from its benchmark index, measured by absolute deviation. A pure index fund has active share zero; a fund with no overlap with its benchmark has active share one hundred. Cremers and Petajisto (2009) showed across the population of US active mutual funds that high-active-share funds (above 80) outperformed their benchmarks both before and after expenses, while closet indexers (active share below 60) underperformed. The unit of attribution is the fund, and by extension the manager who sets its holdings. A high-active-share fund whose realized returns over a short evaluation window are noise-dominated can still be defensibly evaluated using the validated relationship. The evidence is contested: Frazzini, Friedman, and Pomorski (2016) argue that active share's predictive power does not survive benchmark adjustment, and Cremers (2017) responds. We use the example for its structure rather than as a settled empirical claim.

\textbf{Brier scores in geopolitical forecasting.} The Good Judgment Project's tournament population of approximately 25,000 forecasters generated enough resolved-question observations to validate Brier scoring as a predictor of future forecasting accuracy (Mellers et al.~2014). The population-level validation, not any individual forecaster's score in isolation, is what justifies individual-level use of Brier metrics. Tournament structure is also the structural intervention that moved geopolitical forecasting from the unknowable region into detectability at all. The unit of attribution is the individual forecaster.

\textbf{Chief executive trait assessments.} Kaplan, Klebanov, and Sorensen (2012) validated a battery of trait assessments across approximately 300 chief executives, showing that specific trait clusters (execution skills, persistence, analytical orientation) predicted firm performance. Individual chief executive evaluation can then weight these traits using the population-estimated coefficients; the unit of attribution is the individual executive. The history of large-scale trait projects also counsels modesty about adoption: House and colleagues' GLOBE program assembled exactly the kind of cross-context leadership data this methodology calls for (House et al.~2004), and its findings never fully persuaded the field. Building the validation apparatus and earning its acceptance are separate significant challenges.

\textbf{Stage-gate decision quality in pharmaceutical research and development.} The pharmaceutical industry has accumulated decades of stage-gate data across thousands of programs, and the predictive relationships between early-stage decision quality and ultimate program success have been characterized in a large literature. The unit of attribution here deserves care, and illustrates the general point. Stage-gate discipline is most naturally a property of the development process, a refined and domain-specific cousin of the generic management practices Bloom and Van Reenen (2007) measure at the firm level. Attributing it to the head of product development personally requires the further step of showing that the discipline travels with the executive across contexts rather than residing in the organization. Individual program leaders can be evaluated against these benchmarks with statistical defensibility that no individual program's outcome alone could provide, but the default reading is process-level. Pharma has done much of this validation work; finance has done some; venture capital, founder evaluation, and corporate executive selection have done substantially less.

The pattern across domains is consistent. Where validated metrics exist, use them. Where they do not, the framework's prescription includes a research agenda alongside the practical recommendation: commission the population-level studies that would validate them. Pharma is not unique in being amenable to this evaluation methodology; it is simply ahead because the industry has invested more in the validation work. The same methodology can be transplanted to the other unknowable-region domains. The nearest existing large-scale implementation is people analytics: firms now routinely validate selection and performance features across thousands of lower and middle managers, where per-actor observation counts are high and roles are comparable. The irony the map exposes is that this machinery thins out exactly where the stakes concentrate. The positions society most wants to evaluate, the chief executive above all, are the positions that generate the fewest independent outcomes and the least cross-actor comparability.

The methodology also has a structural limit. Population validation requires comparability: a process metric can be validated across hundreds of actors only if it means roughly the same thing for each of them. This requirement necessarily favors generic practices. Bloom and Van Reenen (2007) faced exactly this constraint in measuring management quality across thousands of firms; their instrument necessarily concentrated on operational practices (monitoring, targets, incentives) that apply everywhere, and for that reason it captures the transferable layer of management rather than what strategy scholars mean by strategy, the idiosyncratic use of a firm's idiosyncratic bundle of capabilities and position. The framework's own logic says this limit is not incidental. An idiosyncratic strategic choice is, by construction, an observation without a population; it cannot generate the cross-actor replication that validation requires, any more than a founder's single outcome can. The conclusion is symmetrical: population-validated metrics rescue evaluation of the generic, transferable component of managerial skill, while the idiosyncratic component remains inside the unknowable region. This is not because the methodology is underdeveloped but because idiosyncrasy is the condition of being unvalidatable. Evaluators should draw the line explicitly: credit generic competence on validated metrics, and treat claims of idiosyncratic strategic genius with the same discipline the map applies to outcome records. The success-study genre is the mirror image of this discipline: it begins with winners, codes their practices retrospectively through informants who know the outcomes, and validates nothing out of sample (Rosenzweig 2007a). Population validation inverts each step: it begins with a population that includes the failures, measures features before outcomes resolve, and demonstrates predictive validity or not.

The prescription is straightforward in principle but routinely violated in institutional practice. The single most common mistake is weighting outcomes most heavily where they carry least signal. Boards firing chief executives after three years of poor benchmark-adjusted performance (Jenter and Kanaan 2015), limited partners allocating capital among venture funds on the basis of three vintage records (Lerner, Schoar, and Wongsunwai 2007), and voters crediting and blaming politicians for macroeconomic outcomes the politicians had limited ability to influence (Achen and Bartels 2016) are all cases where the evaluation regime weights outcomes most heavily precisely in the unknowable-region domains. The framework's prediction is that the conventional practice is exactly backwards. A second common mistake is using process metrics that are not population-validated. ``Use process metrics'' without the population-validation discipline is just applying heuristics at a more fine-grained level. A board that decides to evaluate a chief executive on ``decision-making quality'' without an empirical specification of what decision-making features predict outcomes across the population of chief executives is substituting one ungrounded judgment for another.

Active share, trait clusters, and stage-gate quality are not randomly assigned to funds, executives, and programs, so the estimated feature-to-outcome relationships are exposed to omitted variables and selection. Funds that choose high active share may differ in ways that independently produce returns, and executives who score well on execution traits may be hired into firms already positioned to perform. Randomized assignment of the features is rarely feasible, and the medical analogy carries the same caveat, since Framingham itself is an observational cohort whose risk factors required decades of triangulation, and in some cases randomized trials, before being treated as causal. For the evaluator's purpose the limitation is real but not necessarily disqualifying, because evaluation requires stable prediction rather than causal identification. A feature that reliably predicts outcomes across a population carries inferential weight about an individual even if the mechanism is partly confounded. The causal question returns the moment the metric is used prescriptively, to change behavior rather than to evaluate it, and it compounds the gaming problem noted below, since a confounded predictor is precisely the kind that breaks down when actors begin to target it. Two further qualifications are associated with any population-estimated relationship. The estimate is an average effect, so applying it to an individual assumes the actor is exchangeable with the validation population, and this is less justified as the actor or context becomes more distinctive. And the mapping from features to outcomes must be stable. A relationship validated in one industry, era, or regime carries diminishing weight as the deployment context drifts from the validation context.

The prescription is integration rather than substitution. The population-validated process metric supplies an empirically grounded prior on the actor's skill; the noisy outcome record supplies a likelihood; and the evaluator's posterior combines the two with weights set by their reliabilities. The knowability map determines the weight on the outcome side: as \(\text{SNR}^2 \times N_{\text{eff}}\) grows, the record earns weight, and in the essentially unknowable region the posterior is dominated by the process prior, which is what substitution means in practice. This framing prevents two errors, discarding outcome information entirely in domains where it retains modest weight, and treating process metrics as a replacement oracle rather than as one noisy input whose reliability was itself estimated on a population.

\textbf{Structural advantage from prior outcomes.} A second correction to outcome-based attribution complicates the picture in domains where outcomes do carry signal. Even when the framework or a population-validated process metric detects an actor as exceptional, the apparent skill is typically contaminated by structural advantages that arose from the actor's prior outcomes. Better deals come to top venture firms; better recruits come to top coaches; more capital comes to top managers. These are not luck-driven advantages in the simple sense; they are advantages that the actor's prior performance generated and that subsequent performance now reflects. The framework's mean-shift analysis cannot disentangle the portion of an apparent effect attributable to intrinsic skill from the portion attributable to cumulative advantage generated by prior performance. Sorensen (2007) shows quantitatively that approximately two-thirds of the apparent persistence in top-quartile venture-capital returns reflects selection of deal flow rather than investment skill per se. Similar mechanisms operate in chief executive performance (Bertrand and Mullainathan 2001) and across the broader cumulative-advantage literature (Merton 1968; Denrell 2003; Salganik, Dodds, and Watts 2006). Denrell (2003) adds the sampling version of the mechanism: because failures exit the observable population, the survivors available for study carry inflated apparent skill, and the practices common among survivors get mistaken for causes of their survival. The corrections are therefore required in practice: in unknowable domains, substitute population-validated process metrics; in domains where outcomes do carry signal, distinguish total advantage from intrinsic skill before crediting the actor with the latter.

Process metrics are themselves noisy. A board that evaluates a chief executive on decision-quality audits is exposed to the noise and bias of the auditors. The framework does not claim that process metrics are noise-free; it claims only that they are less noise-dominated than the outcome record in essentially unknowable domains, especially when the metrics have been population-validated. Process metrics can also be gamed: a chief executive who knows that the board is auditing decision quality may invest in producing impressive-looking decision narratives rather than in making good decisions. The Goodhart-Lucas problem applies in full. The framework does not solve it, but it shifts the gaming target away from outcome metrics that are equally gameable and substantially less informative.

\section{Conclusion}

The chess grandmaster is genuinely and observably better than the club player. The framework agrees, and tells us why we can be so confident. The hedge fund manager who beat the market for five years may also be genuinely better than one who did not, but we cannot know that from the record alone.

When outcome records carry enough signal to support reliable individual-level inference about skill is determined by the position of the domain on a two-dimensional knowability map defined by per-observation signal-to-noise ratio and effective sample size over the observation window. The detectability inequality SNR \(\times \sqrt{N_{\text{eff}}} > 2.5/d\) tells the evaluator, for any specified effect size \(d\), whether the inference they want to draw is supportable. When outcome records do not carry enough signal, evaluators should substitute population-validated process and structural metrics, following the same methodology medicine uses for individual risk prediction.

The map's most uncomfortable feature is that the domains in which capital, prestige, and political power are most often allocated on the basis of performance attributions---mutual fund management, venture capital, and executive performance---sit firmly in the essentially unknowable region. The framework predicts that the conventional practice of attributing observed performance to skill in these domains is not supportable from outcome data alone, except possibly for actors at the very top tails of their distributions. Even there, apparent skill conflates intrinsic transferable ability with the cumulative structural advantages that prior outcomes generated.

Boards routinely fire chief executives on records the framework predicts cannot support reliable inference. Limited partners routinely allocate capital among venture funds on records that the framework predicts conflate skill with deal-flow advantage. The distinction is consequential for the allocator because deal-flow advantage may attach to the franchise rather than to the partners, so it need not survive the partner departures, successions, and fund-size expansions that are precisely the occasions on which reinvestment decisions are made. Paying premium terms for franchise gravity may still be rational, but it is a different bet than paying for transferable investment skill. Aspiring founders look to the biographies of successful founders for lessons the framework predicts cannot be extracted from any single career. These practices may be warranted, but they rest on weaker evidentiary foundations than their practitioners assume.

\section{References}

Achen, Christopher H., and Larry M. Bartels. 2016. \emph{Democracy for Realists: Why Elections Do Not Produce Responsive Government.} Princeton, NJ: Princeton University Press.

Barras, Laurent, Olivier Scaillet, and Russ Wermers. 2010. ``False Discoveries in Mutual Fund Performance: Measuring Luck in Estimated Alphas.'' \emph{Journal of Finance} 65 (1): 179--216.

Berk, Jonathan B., and Richard C. Green. 2004. ``Mutual Fund Flows and Performance in Rational Markets.'' \emph{Journal of Political Economy} 112 (6): 1269--1295.

Bertrand, Marianne, and Sendhil Mullainathan. 2001. ``Are CEOs Rewarded for Luck? The Ones Without Principals Are.'' \emph{Quarterly Journal of Economics} 116 (3): 901--932.

Bertrand, Marianne, and Antoinette Schoar. 2003. ``Managing with Style: The Effect of Managers on Firm Policies.'' \emph{Quarterly Journal of Economics} 118 (4): 1169--1208.

Bloom, Nicholas, and John Van Reenen. 2007. ``Measuring and Explaining Management Practices Across Firms and Countries.'' \emph{Quarterly Journal of Economics} 122 (4): 1351--1408.

Cochrane, John H. 2005. ``The Risk and Return of Venture Capital.'' \emph{Journal of Financial Economics} 75 (1): 3--52.

Cohen, Jacob. 1988. \emph{Statistical Power Analysis for the Behavioral Sciences.} 2nd ed.~Hillsdale, NJ: Lawrence Erlbaum Associates.

Collins, Jim. 2001. \emph{Good to Great: Why Some Companies Make the Leap\ldots{} and Others Don't.} New York: HarperBusiness.

Cremers, K. J. Martijn. 2017. ``Active Share and the Three Pillars of Active Management: Skill, Conviction, and Opportunity.'' \emph{Financial Analysts Journal} 73 (2): 61--79.

Cremers, K. J. Martijn, and Antti Petajisto. 2009. ``How Active Is Your Fund Manager? A New Measure That Predicts Performance.'' \emph{Review of Financial Studies} 22 (9): 3329--3365.

Denrell, Jerker. 2003. ``Vicarious Learning, Undersampling of Failure, and the Myths of Management.'' \emph{Organization Science} 14 (3): 227--243.

Denrell, Jerker. 2004. ``Random Walks and Sustained Competitive Advantage.'' \emph{Management Science} 50 (7): 922--934.

Denrell, Jerker, Christina Fang, and Daniel Levinthal. 2004. ``From T-Mazes to Labyrinths: Learning from Model-Based Feedback.'' \emph{Management Science} 50 (10): 1366--1378.

Denrell, Jerker, and Chengwei Liu. 2012. ``Top Performers Are Not the Most Impressive When Extreme Performance Indicates Unreliability.'' \emph{Proceedings of the National Academy of Sciences} 109 (24): 9331--9336.

Denrell, Jerker, and James G. March. 2001. ``Adaptation as Information Restriction: The Hot Stove Effect.'' \emph{Organization Science} 12 (5): 523--538.

De Vany, Arthur. 2003. \emph{Hollywood Economics: How Extreme Uncertainty Shapes the Film Industry.} London: Routledge.

Elo, Arpad E. 1978. \emph{The Rating of Chessplayers, Past and Present.} New York: Arco Publishing.

Fama, Eugene F., and Kenneth R. French. 2010. ``Luck versus Skill in the Cross-Section of Mutual Fund Returns.'' \emph{Journal of Finance} 65 (5): 1915--1947.

Frank, Robert H. 2016. \emph{Success and Luck: Good Fortune and the Myth of Meritocracy.} Princeton, NJ: Princeton University Press.

Frazzini, Andrea, Jacques Friedman, and Lukasz Pomorski. 2016. ``Deactivating Active Share.'' \emph{Financial Analysts Journal} 72 (2): 14--21.

Galton, Francis. 1886. ``Regression Towards Mediocrity in Hereditary Stature.'' \emph{Journal of the Anthropological Institute of Great Britain and Ireland} 15: 246--263.

Gilovich, Thomas, Robert Vallone, and Amos Tversky. 1985. ``The Hot Hand in Basketball: On the Misperception of Random Sequences.'' \emph{Cognitive Psychology} 17 (3): 295--314.

Gompers, Paul, Anna Kovner, Josh Lerner, and David Scharfstein. 2010. ``Performance Persistence in Entrepreneurship.'' \emph{Journal of Financial Economics} 96 (1): 18--32.

House, Robert J., Paul J. Hanges, Mansour Javidan, Peter W. Dorfman, and Vipin Gupta, eds.~2004. \emph{Culture, Leadership, and Organizations: The GLOBE Study of 62 Societies.} Thousand Oaks, CA: Sage Publications.

Jenter, Dirk, and Fadi Kanaan. 2015. ``CEO Turnover and Relative Performance Evaluation.'' \emph{Journal of Finance} 70 (5): 2155--2184.

Kahneman, Daniel, and Amos Tversky. 1973. ``On the Psychology of Prediction.'' \emph{Psychological Review} 80 (4): 237--251.

Kaplan, Steven N., Mark M. Klebanov, and Morten Sorensen. 2012. ``Which CEO Characteristics and Abilities Matter?'' \emph{Journal of Finance} 67 (3): 973--1007.

Knight, Frank H. 1921. \emph{Risk, Uncertainty, and Profit.} Boston: Houghton Mifflin.

Korteweg, Arthur G., and Morten Sorensen. 2017. ``Skill and Luck in Private Equity Performance.'' \emph{Journal of Financial Economics} 124 (3): 535--562.

Lerner, Josh, Antoinette Schoar, and Wan Wongsunwai. 2007. ``Smart Institutions, Foolish Choices: The Limited Partner Performance Puzzle.'' \emph{Journal of Finance} 62 (2): 731--764.

Liu, Chengwei, and Mark de Rond. 2016. ``Good Night, and Good Luck: Perspectives on Luck in Management Scholarship.'' \emph{Academy of Management Annals} 10 (1): 409--451.

Loungani, Prakash. 2001. ``How Accurate Are Private Sector Forecasts? Cross-Country Evidence from Consensus Forecasts of Output Growth.'' \emph{International Journal of Forecasting} 17 (3): 419--432.

March, James C., and James G. March. 1977. ``Almost Random Careers: The Wisconsin School Superintendency, 1940--1972.'' \emph{Administrative Science Quarterly} 22 (3): 377--409.

Mauboussin, Michael J. 2012. \emph{The Success Equation: Untangling Skill and Luck in Business, Sports, and Investing.} Boston: Harvard Business Review Press.

Mellers, Barbara, Lyle Ungar, Jonathan Baron, Jaime Ramos, Burcu Gurcay, Katrina Fincher, Sydney E. Scott, Don Moore, Pavel Atanasov, Samuel A. Swift, Terry Murray, Eric Stone, and Philip E. Tetlock. 2014. ``Psychological Strategies for Winning a Geopolitical Forecasting Tournament.'' \emph{Psychological Science} 25 (5): 1106--1115.

Mellers, Barbara, Eric Stone, Terry Murray, Angela Minster, Nick Rohrbaugh, Michael Bishop, Eva Chen, Joshua Baker, Yuan Hou, Michael Horowitz, Lyle Ungar, and Philip Tetlock. 2015. ``Identifying and Cultivating Superforecasters as a Method of Improving Probabilistic Predictions.'' \emph{Perspectives on Psychological Science} 10 (3): 267--281.

Merton, Robert K. 1968. ``The Matthew Effect in Science.'' \emph{Science} 159 (3810): 56--63.

Rosenzweig, Phil. 2007a. \emph{The Halo Effect\ldots{} and the Eight Other Business Delusions That Deceive Managers.} New York: Free Press.

Rosenzweig, Phil. 2007b. ``Misunderstanding the Nature of Company Performance: The Halo Effect and Other Business Delusions.'' \emph{California Management Review} 49 (4): 6--20.

Salganik, Matthew J., Peter Sheridan Dodds, and Duncan J. Watts. 2006. ``Experimental Study of Inequality and Unpredictability in an Artificial Cultural Market.'' \emph{Science} 311 (5762): 854--856.

Sorensen, Morten. 2007. ``How Smart Is Smart Money? A Two-Sided Matching Model of Venture Capital.'' \emph{Journal of Finance} 62 (6): 2725--2762.

Tetlock, Philip E. 2005. \emph{Expert Political Judgment: How Good Is It? How Can We Know?} Princeton, NJ: Princeton University Press.

Tversky, Amos, and Daniel Kahneman. 1971. ``Belief in the Law of Small Numbers.'' \emph{Psychological Bulletin} 76 (2): 105--110.

Tversky, Amos, and Daniel Kahneman. 1974. ``Judgment under Uncertainty: Heuristics and Biases.'' \emph{Science} 185 (4157): 1124--1131.

Wermers, Russ. 2000. ``Mutual Fund Performance: An Empirical Decomposition into Stock-Picking Talent, Style, Transactions Costs, and Expenses.'' \emph{Journal of Finance} 55 (4): 1655--1695.

\end{document}